\documentclass[]{aa}
\usepackage{geometry}
\usepackage{graphicx}
\usepackage{txfonts}
\usepackage{amsmath}
\usepackage{amssymb}
\usepackage{aas_macros}
\usepackage{booktabs}
\usepackage{graphicx}
\usepackage{hyperref}
\usepackage{listings}
\usepackage{multirow}
\usepackage{extarrows}
\usepackage{xcolor}
\usepackage{bm}

\newcommand{\LTE}{\mathrm{LTE}}
\newcommand{\LDTE}{\mathrm{LDTE}}

\newcommand{\TCB}{\mathrm{TCB}}
\newcommand{\TDB}{\mathrm{TDB}}

\newcommand{\TCL}{\mathrm{TCL}}

\newcommand{\M}{\mathrm{M}}

\newcommand{\LB}{L_{B}}

\newcommand{\kbo}{Kuiper belt object}
\newcommand{\te}{time ephemeris}
\newcommand{\tes}{time ephemerides}
\newcommand{\lte}{lunar time ephemeris}
\newcommand{\ltes}{lunar time ephemerides}

\begin{document}
   \title{Lunar Time Ephemeris \texttt{LTE440}: definitions, algorithm and performance }



   \author{Xu Lu \inst{1}
        \and Tian-Ning Yang \inst{1,2}
        \and Yi Xie \inst{1}
        }

   \institute{Purple Mountain Observatory, Chinese Academy of Sciences, Nanjing 210023, China\\
   \email{yixie@pmo.ac.cn}
            \and School of Astronomy and Space Science, University of Science and Technology of China, Hefei 230026, China\\ }

   \date{Received September 30, 20XX}

 
  \abstract
  {Robotic and human activities in the cislunar space are expected to rapidly increase in the future. 
  Modeling, jointly analysis and sharing of time measurements made in the vicinity of the Moon might indispensably demand calculating a lunar time scale and transforming it into other time scales.}
  {For users, we present a ready-to-use software package of Lunar Time Ephemeris \texttt{LTE440} that can calculate the Lunar Coordinate Time (TCL) and its relations with the Barycentric Coordinate Time (TCB) and the Barycentric Dynamical Time (TDB).} 
   {According to the International Astronomical Union Resolutions on relativistic time scales, we numerically calculate the relativistic time-dilation integral in the transformation between TCL and TCB/TDB with the JPL ephemeris DE440 including the gravitational contributions from the Sun, all planets, the main belt asteroids and the Kuiper belt objects, and export data files in the SPICE format.  }
  {At a conservative estimate, \texttt{LTE440} has an accuracy better than 0.15 ns before 2050 and a numerical precision at the level of 1 ps over its entire time span. 
  The secular drifts between the coordinate times in \texttt{LTE440} are respectively estimated as 
  $\langle \mathrm{d}\,\mathrm{TCL}/\mathrm{d}\,\mathrm{TCB}\rangle=1-1.482\,536\,216\,7\times10^{-8}$
  and
  $\langle \mathrm{d}\,\mathrm{TCL}/\mathrm{d}\,\mathrm{TDB}\rangle=1+6.798\,355\,24\times10^{-10}$.
  Its most significant periodic variations are an annual term with amplitude of 1.65 ms and a monthly term with amplitude of 126 $\muup$s.}
  {\texttt{LTE440} might satisfy most of current needs and is publicly available.}

   \keywords{time --
                ephemerides
               }

   \maketitle


\nolinenumbers

\section{Introduction}


Robotic and human activities in the cislunar space are expected to rapidly increase in the future. Modeling, jointly analysis and sharing of time measurements made in the vicinity of the Moon might indispensably demand calculating a lunar time scale and transforming it into other time scales. For users, we present a ready-to-use software package of Lunar Time Ephemeris $\texttt{LTE440}$ that can calculate the Lunar Coordinate Time (TCL) and its relations with the Barycentric Coordinate Time (TCB) and the Barycentric Dynamical Time (TDB). According to the International Astronomical Union Resolutions on relativistic time scales, we numerically calculate the relativistic time-dilation integral in the transformation between TCL and TCB/TDB with the JPL ephemeris DE440 including the gravitational contributions from the Sun, all planets, the main belt asteroids and the Kuiper belt objects, and export data files in the SPICE format.  At a conservative estimate, \texttt{LTE440} has an accuracy better than 0.15 ns before 2050 and a numerical precision at the level of 1 ps over its entire time span. The secular drifts between the coordinate times in \texttt{LTE440} are respectively estimated as $\langle \mathrm{d}\,\mathrm{TCL}/\mathrm{d}\,\mathrm{TCB}\rangle=1-1.482\,536\,216\,7\times10^{-8}$ and $\langle \mathrm{d}\,\mathrm{TCL}/\mathrm{d}\,\mathrm{TDB}\rangle=1+6.798\,355\,24\times10^{-10}$. Its most significant periodic variations are an annual term with amplitude of 1.65 ms and a monthly term with amplitude of 126 $\mu$s. \texttt{LTE440} might satisfy most of current needs and is publicly available.

The relativistic time-dilation integral is an indispensable ingredient in the transformation between time scales those are defined in two distinct reference systems in the Solar System.

One perfect example is the \te\ of the Earth, traditionally called ``\te'' for short, 
that is a representation of the integral in the transformation between
the Geocentric Coordinate Time (TCG)/Terrestrial Time (TT) and
the Barycentric Coordinate Time (TCB)/Barycentric Dynamical Time (TDB), according to the International Astronomical Union (IAU) Resolutions of relativistic reference systems and time scales \citep{Soffel2003AJ126.2687,IAU2006Res.B3}. The \te\ has been an essential part of the current framework for the reference systems of the relativistic spacetime \citep[e.g.][]{Fukushima1996IAUS172.461,Brumberg1996CMDA64.231,Fukushima2001cemesymp.230,Seidelmann2002A&A392.341,Bretagnon2003A&A408.387,Soffel2003AJ126.2687,Kopeikin2007AIPC886.268,Kopeikin2010IAUS261.7,Turyshev2025ApJ985.140,Kaplan2005USNOC179,Klioner2008A&A478.951,McCarthy2011Metrologia48.S132}
and has been widely used in Earth-based time-related measurements, such as precise modeling of astrometry \citep[e.g.][]{Klioner2003AJ125.1580,Lindegren2003A&A401.1185,Fukushima2005ASPC338.343}, 
spacecraft tracking \citep[e.g.][]{Turyshev1996arXiv9606.028,Sekido2006JGeod80.137,Dirkx2016A&A587.A156}, 
pulsar timing \citep[e.g.][]{Kopeikin1997PRD56.4455,Kopeikin1999MNRAS305.563,Hobbs2006MNRAS369.655,Edwards2006MNRAS372.1549}
and geodesy \citep{Muller2008JGeod82.133,Sanchez2016JGeod90.815}. 
The \tes\ can be divided into two kinds: analytical and numerical. The analytical ones might trace back to those in 1960s \citep{Aoki1964AJ69.221,Clemence1967AJ72.1324}, develop to those more accurate and complex \citep{Moyer1981CM23.33,Moyer1981CM23.57,Hirayama1987TSCM.75}
and climax with \citet{Fairhead1990A&A229.240}, all of which are based on analytical planetary and lunar ephemerides. The numerical \te\ evaluates the relativistic time-dilation integral numerically with the quantities provided by a numerical planetary and lunar ephemeris. It starts with \citet{Backer1986ARA&A24.537} and evolves to the works of  \citet{Fukushima1995A&A294.895} and \citet{Irwin1999A&A348.642} with more sophisticated method and higher accuracy. Nowadays, all of the most widely-used planetary and lunar ephemerides, such as DE440 \citep{Park2021AJ161.105}, INPOP21a \citep{Fienga2021NSTIMC110} and EPM2021 \citep{Pitjeva2022IAUS364.220}, provide evaluated TT$-$TDB in their products which can be directly taken by a user without any further calculation and INPOP21a also gives TCG$-$TCB.

However, with the expected increase of robotic and human activities in the cislunar space in the future, new demands might emerge.
A \lte, that represents the evaluation of the relativistic time-dilation in the transformation between the Lunar Coordinate Time (TCL) \citep{IAU2024ResII} and TCB/TDB, 
would be required for those
who would like to conveniently model, jointly analyse and share their time measurements made in the vicinity of the Moon.
A user might take advantage of the \lte' high accuracy and ease of use in certain scenarios, including (but not limited to) tracing back the proper time of a Moon-based clock to the Coordinated Universal Time (UTC) and conducting the very long baseline interferometry (VLBI) with telescopes separated in the neighbours of the Earth and Moon.


Missions on and around the Moon in the future might require a practical lunar reference time (LRT) \citep{IAU2024ResIII}, 
since TCL, defined as the time coordinate for the Lunar Celestial Reference System (LCRS) \citep{IAU2024ResII}, is not given by any real clock.
Although international organizations and countries have not achieved any agreement on the definition of such a reference time scale for now,
all agree that the relationship between LRT and UTC must be clear.
This traceability demands to be bridged by a \lte. 
If the uncertainty of LRT$-$UTC is supposed to be within 10 ns in a day, it translates to a relative error and corresponding \lte\ derivative error of $10^{-13}$.
The systematic errors of the \lte\ and its derivative should be ideally kept at the level of 0.1 ns and $10^{-15}$, at least 2 orders of magnitude smaller than proposed demands,
according to the consideration of \citet{Irwin1999A&A348.642}.

By taking advantage of low gravity, absence of atmosphere, deep natural cooling, low magnetic field and interference-free radio environment of the Moon \citep{Burns1985LBSAConf.293,Gurvits1998HoA11B.985},
Earth-Moon VLBI with Earth- and Moon-based radio telescopes might significantly improve our understanding of the Moon’s orbital and rotational motion and the precision of astrometry with its longer baselines and more sufficient $uv$-coverage \citep{Kurdubov2019MNRAS486.815}.
With the expectation that Earth-Moon VLBI would reach its best performance at least as good as ground-based VLBI, 
it needs time tags of the recorded signals on the Moon with uncertainty less than 1 $\muup$s
and Moon-based clocks with stability better than $10^{-14}$ at 50 min
based on the knowledge of ground-based VLBI \citep{Nothnagel2018SSR214.66}.
Therefore, the systematic errors of the \lte\ and its derivative need be controlled at the level of 10 ns and $10^{-16}$, 2 orders of magnitude better than these requirements.
Targeted at direct imaging supermassive black holes and their photon rings \citep{Fish2020ASR65.821,Roelofs2021A&A650A.56}, future space-borne sub-millimeter VLBI in the cislunar space might raise even more challenges to the technological and theoretical aspects of time-related measurement \citep{Gurvits2022AcAst196.314,Hudson2023AcAst213.681}.


Very recently, \citet{Ashby2024AJ168.112}, \citet{Kopeikin2024PRD110.084047} and \citet{Turyshev2025ApJ985.140} estimated the rate of an ideal clock on the Moon's selenoid with respect to the one on the Earth's geoid.
Furthermore, \citet{Kopeikin2024PRD110.084047} and \citet{Turyshev2025ApJ985.140} also constructed analytical formulae to handle the relativistic time-dilation integral associated with TCL by adopting different approximations and cut-off for the lunar motion, which could be somehow regarded as fledgling analytic \ltes.  
Motivated by the potential needs and inspired by these pioneer works, we will extend these analytical results by making a numerical \lte\ following the methodology of foundational works by \citet{Fukushima1995A&A294.895} and \citet{Irwin1999A&A348.642}.

In this work, we present the \lte\ package \texttt{LTE440} 
\footnote{\texttt{LTE440} including its data files and codes for usage is publicly available at \url{https://github.com/xlucn/LTE440}.}
that contains numerically evaluation of the relativistic time-dilation integral between TCL and TCB/TDB and also provides functions to calculate TCL$-$TCB, TCL$-$TDB and TCL for a given TDB moment.
We define the relativistic time-dilation integral of the \lte\ in Sec. \ref{sec:def}.
We explain the algorithms for calculating the integral based on DE440 and making the product of \texttt{LTE440} in Sec. \ref{sec:alg}.
By comparing \texttt{LTE440} with \texttt{LTE430} and \texttt{LTE441}, two auxiliary lunar time ephemerides packages based on DE430 and DE441, we analyse their characteristics and differences, assess their performance, and correct the linear drift of
$\langle \mathrm{d}\mathrm{TCL}/\mathrm{d}\mathrm{TCB}\rangle$
of \texttt{LTE440} through \texttt{LTE441} in Sec. \ref{sec:perf}.
We conclude the paper in Sec. \ref{sec:con}.

\section{Definitions}

\label{sec:def}

According to the IAU 2024 Resolution II \citep{IAU2024ResII}, LCRS and TCL can be constructed with the same techniques used to construct the Geocentric Celestial Reference System
(GCRS) and TCG. 
The transformation between TCL and TCB is analogous to the transformation between TCG and TCB given by IAU 2000 Resolution B1.5 \citep{Soffel2003AJ126.2687}, 
with the Earth-related quantities replaced by the Moon-related quantities. 
Therefore, the relation between TCL and TCB can be expressed as
\begin{eqnarray}
\label{eq:TCL-TCB}
  & & \TCL-\TCB \nonumber\\
  & \xlongequal{\TCB} & - \frac{1}{c^2}\int_{t_0}^t\left(\frac{v_\M^2}{2}+w_\mathrm{0M}+w_{l\M}\right)\mathrm{d}t - \frac{1}{c^2}\boldsymbol{v}_\M\cdot(\boldsymbol{x}-\boldsymbol{x}_{\M}) \nonumber \\
  & & +\frac{1}{c^4}\int_{t_0}^t\left(-\frac{v_\M^4}8-\frac{3}{2}v_\M^2w_\mathrm{0M} +  4\boldsymbol{v}_\M\cdot\boldsymbol{w}_{\M} +\frac12w_\mathrm{0M}^2 \right.\nonumber\\
  & & \left.+\Delta_\M\right)\mathrm{d}t - \frac{1}{c^4}\left(3w_\mathrm{0M}+\frac{v_\M^2}2\right)\boldsymbol{v}_\M\cdot(\boldsymbol{x}-\boldsymbol{x}_{\M}),
\end{eqnarray}
where ``$\xlongequal{\TCB}$'' means that all quantities on the its right-hand side are TCB-compatible, 
$t$ is TCB, 
$t_0$=1977 January 1, 0$^\mathrm{h}$0$^\mathrm{m}$32.184$^\mathrm{s}$ is the common value of TT, TCG, TCB and TCL set on 1977 January 1, 0$^\mathrm{h}$0$^\mathrm{m}$0$^\mathrm{s}$ TAI \citep{IAU2024ResII},
$\boldsymbol{x}$ is the barycentric position vectors of the point where the transformation between TCL and TCB is done, $\boldsymbol{x}_{\M}$ and $\boldsymbol{v}_\M$ are the barycentric position and velocity vectors of the Moon. 
The potential $w_\mathrm{0M}$ describes the gravitational potential of external bodies at the center of the Moon in the point-mass approximation and it is defined as
\begin{equation}
  w_\mathrm{0M}=\sum\limits_\mathrm{A\neq M}\frac{GM_\mathrm{A}}{r_\mathrm{MA}},
\end{equation}
where $r_\mathrm{MA}=|\bm{r}_\mathrm{MA}|$,
$\boldsymbol{r}_\mathrm{MA}=\boldsymbol{x}_{\M}-\boldsymbol{x}_\mathrm{A}$, 
and the summation runs over all of the Solar System bodies A except the Moon (M). 
The potential $w_{l\M}$ represents the non-spherical gravitational potential of external bodies. 
The vector potential $\bm{w}_{\M}$ can be expressed as
\begin{equation}
  \bm{w}_{\M}=\sum\limits_\mathrm{A\neq M}G\left[\frac{M_\mathrm{A}\boldsymbol{v}_\mathrm{A}}{r_\mathrm{MA}}-\frac{\boldsymbol{r}_\mathrm{MA}\times\boldsymbol{S}_\mathrm{A}}{2r_\mathrm{MA}^3}\right],
\end{equation}
where $\bm{S}_\mathrm{A}$ is the total angular momentum of body A. 
The $\Delta_\M$ term contains nonlinear couplings of $w_\mathrm{0M}$ and $\bm{w}_{\M}$ and other higher-order contributions at $\mathcal{O}(c^{-4})$ and it is defined as
\begin{eqnarray}
  \Delta_\M&=&\sum\limits_\mathrm{A\neq M}\left\{\frac{GM_\mathrm{A}}{r_\mathrm{MA}}\left[\sum_{\mathrm{B}\neq \mathrm{A}}\frac{GM_\mathrm{B}}{|\boldsymbol{x}_\mathrm{B}-\boldsymbol{x}_\mathrm{A}|}-2v_\mathrm{A}^2+\frac{\left(\boldsymbol{r}_\mathrm{MA}\cdot\boldsymbol{v}_\mathrm{A}\right)^2}{2r_\mathrm{MA}^2}\right.\right.\nonumber\\
  & & + \left.\left.\frac{\boldsymbol{r}_\mathrm{MA}\cdot\boldsymbol{a}_\mathrm{A}}{2}\right]+\frac{2G\boldsymbol{v}_\mathrm{A}\cdot(\boldsymbol{r}_\mathrm{MA}\times\boldsymbol{S}_\mathrm{A})}{r_\mathrm{MA}^3}\right\},
\end{eqnarray}
where $\boldsymbol{a}_\mathrm{A}$ is the barycentric acceleration vector of the body A.

In order to evaluate the relation of TCL$-$TCB \eqref{eq:TCL-TCB}, 
one inevitably needs an ephemeris to know the positions and velocities of the Solar System bodies up to a sufficient accuracy. Given the fact that all the modern numerical ephemerides adopt TDB as their main time scale, it would be more convenient for calculation and utility to rewrite TCL$-$TCB as TCL$-$TDB by considering the linear relation between TCB and TDB \citep{IAU2006Res.B3}
\begin{equation}
\label{eq:TDBdef}
	\TDB=\TCB-L_{B}(\TCB-t_0)+\TDB_0
\end{equation}
so that we find  
\begin{eqnarray}
\label{eq:TCL-TDB}
  &&\TCL-\TDB\nonumber\\
  & \xlongequal{\TDB} &\frac{\LB}{1-\LB}(\TDB-t_0)-\frac{1}{1-\LB}\TDB_0 \nonumber \\
  &&-\frac{1}{1-\LB}\Bigg[\frac{1}{c^2}\int_{\TDB_0+t_0}^\TDB\left(\frac{v_\M^2}{2}+w_\mathrm{0M}+w_{l\M}\right)\mathrm{d}\TDB\nonumber \\
  &&+\frac{1}{c^2}\boldsymbol{v}_\M\cdot(\boldsymbol{x}-\boldsymbol{x}_{\M})-\frac{1}{c^4}\int_{\TDB_0+t_0}^\TDB\left(-\frac{v_\M^4}8-\frac{3}{2}v_\M^2w_\mathrm{0M}\right.\nonumber\\
  &&+\left.4\boldsymbol{v}_\M\cdot\boldsymbol{w}_{\M}+\frac12w_\mathrm{0M}^2+\Delta_\M\right)\mathrm{d}\TDB\nonumber \\
  &&+\frac{1}{c^4}\left(3w_\mathrm{0M}+\frac{v_\M^2}2\right)\boldsymbol{v}_\M\cdot(\boldsymbol{x}-\boldsymbol{x}_{\M})\Bigg],
\end{eqnarray}
where ``$\xlongequal{\TDB}$'' means that all quantities on the its right-hand side are TDB-compatible,
and $L_\mathrm{B}=1.550519768\times10^{-8}$ and $\TDB_0=-6.55\times10^{-5}$ s are defining constants.
If a user can directly combine TCL$-$TDB and TT$-$TDB, the one can easily find the TCL$-$TT. 
Since TT$-$TDB has been very precisely provided by the ephemerides DE440, INPOP21a and EPM2021, we believe a ready-to-use product of TCL$-$TDB may help to build an easy way for tracing back the Moon-based time scales to the Earth-based ones.

Following \citet{Fukushima1995A&A294.895}, we denotes the time-dilation integral in Eq. \eqref{eq:TCL-TCB} as the \lte\ (LTE). 
Considering TDB as the time scale used by modern planetary ephemerides
and following the note on computation of IAU 2000 Resolution B1.5 \citep{Soffel2003AJ126.2687},
we write it in terms of TDB-compatible quantities as
\begin{eqnarray}
  \label{eq:LTE}
  & & \LTE(\TDB) \nonumber\\
  & = & -\frac{1}{1-\LB}\Bigg[\frac{1}{c^2}\int_{\TDB_0+t_0}^\TDB\left(\frac{v_\M^2}{2}+w_\mathrm{0M}+w_{l\M}\right)\mathrm{d}\TDB\nonumber \\
  & & -\frac{1}{c^4}\int_{\TDB_0+t_0}^\TDB\left(-\frac{v_\M^4}8-\frac{3}{2}v_\M^2w_\mathrm{0M}+4\boldsymbol{v}_\M\cdot\boldsymbol{w}_{\M}\right.\nonumber\\
  & & +\left.\frac12w_\mathrm{0M}^2+\Delta_\M\right)\mathrm{d}\TDB\Bigg],
\end{eqnarray}
which is nothing but the integrals in Eq. \eqref{eq:TCL-TDB}.
With its help, we might rewrite the transform of TCL$-$TCB as
\begin{equation}
\label{eq:TCL-TCB=LTE}
    \TCL-\TCB = \LTE(\TDB)+\mathrm{LPB}(\TCB,\bm{x})
\end{equation}
where $\mathrm{LPB}(\TCB,\bm{x})$ depends on TCB and the user's TCB-compatible position as
\begin{eqnarray}
    \mathrm{LPB}(\TCB,\bm{x}) & \xlongequal{\TCB} & - \frac{1}{c^2}\boldsymbol{v}_\M\cdot(\boldsymbol{x}-\boldsymbol{x}_{\M}) \nonumber \\
    && - \frac{1}{c^4}\left(3w_\mathrm{0M}+\frac{v_\M^2}2\right)\boldsymbol{v}_\M\cdot(\boldsymbol{x}-\boldsymbol{x}_{\M}),
\end{eqnarray}
and we might also change the form of TCL$-$TDB into
\begin{equation}
\label{eq:TCL-TDB=LTE}
    \TCL-\TDB  = \LDTE(\TDB)+\mathrm{LPD}(\TDB,\bm{x})
\end{equation}
where
\begin{eqnarray}
\label{eq:LDTE}
    \LDTE(\TDB) & = & \frac{\LB}{1-\LB}(\TDB-t_0)-\frac{1}{1-\LB}\TDB_0 \nonumber\\
    & & +\LTE(\TDB),
\end{eqnarray}
and $\mathrm{LPD}(\TDB,\bm{x})$ depends on TDB and the user's TDB-compatible position as
\begin{eqnarray}
    \mathrm{LPD}(\TDB,\bm{x}) & \xlongequal{\TDB} & \frac{1}{1-\LB}\left[- \frac{1}{c^2}\boldsymbol{v}_\M\cdot(\boldsymbol{x}-\boldsymbol{x}_{\M}) \right.\nonumber \\
    && \left.- \frac{1}{c^4}\left(3w_\mathrm{0M}+\frac{v_\M^2}2\right)\boldsymbol{v}_\M\cdot(\boldsymbol{x}-\boldsymbol{x}_{\M})\right].
\end{eqnarray}
We can see that, for the transformation between TCL$-$TCB, 
the introduction of TDB might bring convenience into the computation of its integral in Eq.~\eqref{eq:TCL-TCB} and into making a ready-to-use product of \lte, 
whereas it might cause an inconvenience that the user have to handle three time coordinates emerged in Eq.~\eqref{eq:TCL-TCB=LTE}. 
The transformation between TCL$-$TDB bridged by LDTE(TDB) in Eq.~\eqref{eq:TCL-TDB=LTE} is free of such a dilemma 
so that, with an available product of LDTE(TDB), a user can easily and clearly calculate TCL for a given TDB moment and vice versa.
Therefore, in the following parts of this work, we would adopt the chain of transformations as
\begin{equation}
    \TCL\xleftrightarrow[\LDTE]{\mathrm{Eq.}~\eqref{eq:TCL-TDB=LTE}}\TDB\xleftrightarrow{}\TCB.
\end{equation}

Using JPL ephemeris DE440 \citep{Park2021AJ161.105}, we numerically compute the right-hand side of Eq. \eqref{eq:LTE} and export data files in the SPICE format. 
For a user, we also provide software functions to calculate LTE(TDB) in TCL$-$TDB and LDTE(TDB) in TCL$-$TCB at a given TDB epoch, respectively.
We call the package of these data files and functions as \texttt{LTE440}. 
In order to provide more understanding about it, we also make two auxiliary lunar time ephemeris packages \texttt{LTE430} and \texttt{LTE441} based on JPL ephemerides DE430 \citep{Folkner2014IPNPR196.1} and DE441 \citep{Park2021AJ161.105} for making comparison and assessing uncertainties. 
For later convenience, we would denote the mathematical function LTE(TDB) \eqref{eq:LTE} in \texttt{LTE430}, \texttt{LTE440} and \texttt{LTE441} as LTE$_{430}$, LTE$_{440}$ and LTE$_{441}$ for short, respectively.

\section{Algorithm}
\label{sec:alg}
We evaluate the integral in Eq. \eqref{eq:LTE} using a 10th-order Romberg scheme \citep{Press1992nrfa.book} with half-day integration intervals. 
The integrand is computed by usage of position and velocity data of the Sun, planets and small objects provided by JPL ephemerides.
Since the integrand consists of a constant value plus periodic terms, we would expect that the resulting \lte\ would have a more significant secular drift along with indistinct periodic terms over a long time span.
Since no mature analytic series of \lte\ is currently available, we are unable to use the hybrid method of \cite{Irwin1999A&A348.642} to separate the secular drift and periodic variations for keeping the accuracy of the numerical evaluation of Eq. \eqref{eq:LTE}. 
In order to reveal these periodic terms in the \lte\ more clearly, 
we adopt a method mixed the approaches of \cite{Fukushima1995A&A294.895} and \cite{Irwin1999A&A348.642}. 
First, we subtract a roughly estimated constant, which is an approximated secular drift rate of the \lte, from the integrand of Eq. \eqref{eq:LTE} in the initial evaluation \citep{Irwin1999A&A348.642}.
Then, we iteratively find the secular drift rate of the resulting curve by fitting it with a linear function and re-evaluate the integration after removal of this rate from the integrand until this rate converges \citep{Fukushima1995A&A294.895}.
Finally, we subtracted this converged rate from the integrand, redo the integration numerically and combine the evaluation with the secular drift as our product. 
We suppose that this procedure might be able to more clearly separate the secular drift and the periodic terms in the \lte\ based solely on the numerical computation.
We also discuss a possible way to calibrate such a drift rate by using on a longer time-span numerical ephemeris in Sect. \ref{sec:con}.

We would use three versions of JPL ephemerides DE430, DE440 and DE441 to produce the lunar time ephemerides
and the characteristics of these planetary ephemerides are summarized in Table \ref{tab:dynmod} for a brief comparison.
As pointed by \citet{Park2021AJ161.105}, several updates have been made in DE440/DE441 comparing to DE430, including 30 Kuiper belt objects and a ring, solar radiation pressure on the Earth and Moon, and the Lense–Thirring effect from the Sun's angular momentum.
For the resulting \ltes, gravitational contribution of these \kbo s and the motion of the Moon could be the main factors for making \texttt{LTE430} and \texttt{LTE440}/\texttt{441} different from each other.

\begin{figure}
  \centering
  \includegraphics[width=\hsize]{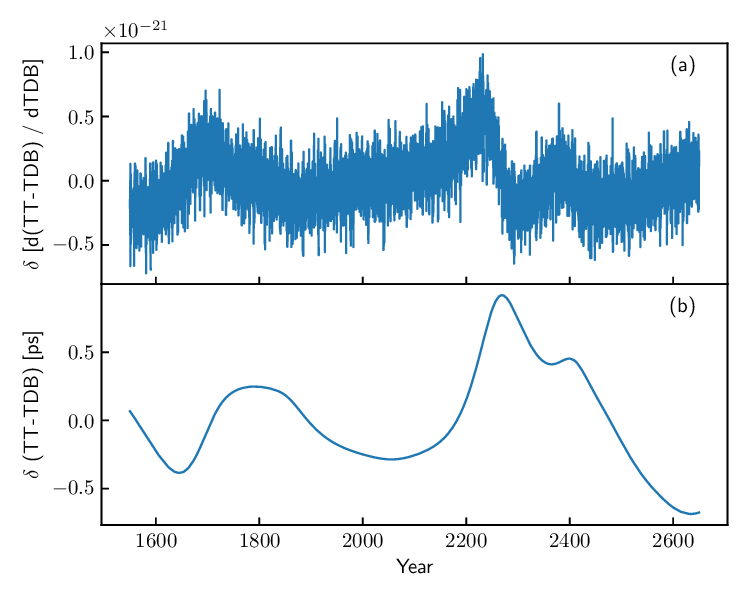}
  \caption{Differences between our reproduced version of TT$-$TDB at the geocenter with the planetary data of DE440 and the inherent version of DE440. Top panel: difference between the derivatives $\delta[\mathrm{d(TT-TDB)}/\mathrm{dTDB}]=[\mathrm{d(TT-TDB)}/\mathrm{dTDB}]_\mathrm{this}-[\mathrm{d(TT-TDB)}/\mathrm{dTDB}]_\mathrm{DE440}$. Bottom panel: difference between the Earth time ephemerides $\delta\mathrm{(TT-TDB)}=\mathrm{(TT-TDB)}_\mathrm{this}-\mathrm{(TT-TDB)}_\mathrm{DE440}$.}\label{fig:ttmtdb_comparison}
\end{figure}

We estimate the error of our integration scheme by comparing the resulting \lte\ with integration intervals of half-day and one-day, respectively.
We find that their difference is at the order of 10 fs, which is negligible for the purpose of this work.
We further assess the accuracy of our numerical method by reproducing TT$-$TDB at the geocenter with planetary data of DE440 and comparing our result with the one inherently provided by DE440.
As shown in Fig. \ref{fig:ttmtdb_comparison}, we find that the difference between our reproduction and the one of DE440 is less than 1 ps over their entire time span,
demonstrating that our numerical scheme is adequate to produce a precise \lte\ comparable with the \te\ provided by DE440.  

The values of the integral \eqref{eq:LTE} with its secular drift removed are fitted with 13th-degree Chebyshev polynomials on 4-day granules using the method described by \cite{Newhall1989CM45.305}. 
The coefficients of these Chebyshev polynomials are exported into a binary SPK file conforming to the SPICE format.
The coefficients of the secular drift are stored in a text PCK file.
These files for \texttt{LTE440} are publicly available online and 
see \cite{Lu2025arXiv2506.19213} for more details of usage.

\section{Performance}
\label{sec:perf}

In this section, we present some detailed characteristics of \texttt{LTE440}, including its secular drift and periodic variations, make a comparison between \texttt{LTE430} and \texttt{LTE440}, and try to correct the secular drift of \texttt{LTE440} by using \texttt{LTE441} with a longer time-span.

\subsection{Characteristics}


It is well known that there are secular drifts of TCG with respect to TCB and to TDB \citep{Seidelmann1992A&A265.833}. It is also true for TCL. Following IAU Resolution 2000 \citep{Soffel2003AJ126.2687}, we introduce $L_{C}^{\M}$ and $L_{D}^{\M}$ to describe the secular drifts of TCL with respect to TCB and TDB, respectively, as
\begin{equation}
  \label{eq:<dTCLdTCB>}
 \left\langle \frac{\mathrm{d}\,\mathrm{TCL}}{\mathrm{d}\,\mathrm{TCB}}\right\rangle = 1-L_{C}^{\M},
\end{equation}
and
\begin{equation}
  \label{eq:<dTCLdTDB>}
 \left\langle \frac{\mathrm{d}\,\mathrm{TCL}}{\mathrm{d}\,\mathrm{TDB}}\right\rangle = 1+L_{D}^{\M},
\end{equation}
where $\langle \  \rangle$ refers to a sufficient long time average to remove its periodic changes. $L_{C}^{\M}$ and $L_{D}^{\M}$ satisfy the following relation
\begin{equation}
    (1-L_{C}^{\M})=(1-L_{B})(1+L_{D}^{\M}).
\end{equation}
In fact, according to Eqs.~\eqref{eq:TDBdef} and \eqref{eq:TCL-TCB=LTE}, we can have that, at the selenocenter $\bm{x}=\bm{x}_{\M}$,
\begin{equation}
    \label{}
    L_{C}^{\M} = - (1-L_B) \left\langle \frac{\mathrm{d}\,\mathrm{LTE}}{\mathrm{d}\,\mathrm{TDB}}\right\rangle.
\end{equation}
Based on \texttt{LTE440}, we find that
\begin{equation}
\label{eq:L_C^MLTE440}
 L_{C,440}^{\M}= 1.482\,536\,216\,7\times10^{-8}\pm1.0\times10^{-17},
\end{equation}
and
\begin{equation}
\label{eq:L_D^MLTE440}
 L_{D,440}^{\M}= 6.798\,355\,24\times10^{-10}\pm1.0\times10^{-17}.
\end{equation}
Our numerical estimation of $L_{C}^{\M}$ is consistent with the analytic ones given by \citet{Kopeikin2024PRD110.084047} and \citet{Turyshev2025ApJ985.140} (see Table \ref{tab:L_C^M} for a comparison).

\begin{table}
\caption{A summary of estimated $L_{C}^{\M}$.}
\label{tab:L_C^M}
\centering
\begin{tabular}{ll}
\hline\hline
$L_{C}^{\M}$\ ($10^{-8}$) & Ref. \\
\hline
$1.482\,536\,214\,9$ & This work, \texttt{LTE430}\\
$1.482\,536\,216\,7$ & This work, \texttt{LTE440}\\
$1.482\,536\,221\,7$ & This work, \texttt{LTE440} calibrated with \texttt{LTE441}\\
$1.482\,4$\tablefootmark{a} & \citet{Kopeikin2024PRD110.084047} \\
$1.482\,536\,24$\tablefootmark{b} & \citet{Turyshev2025ApJ985.140} \\
\hline
\end{tabular}
\tablefoot{\tablefoottext{a}{Derived from the value of $1.280\,8$ ms d$^{-1}$ given by \citet{Kopeikin2024PRD110.084047}}. \tablefoottext{b}{Denoted as $L_\mathrm{H}$ in \citet{Turyshev2025ApJ985.140}}.
}
\end{table}

\begin{table*}
\caption{Contributing sources in $L_{C}^{\M}$ of LTE$_{430}$ and LTE440$_{440}$.}
\label{tab:LC_i} 
\centering
\begin{tabular}{llllr}
\hline\hline             
Term & Gravitational Body & \multicolumn{3}{c}{Contribution to $L_{C}^{\M}$} \\
&& LTE$_{430}$ & LTE$_{440}$& LTE$_{430}$$-$LTE$_{440}$\\
\hline
 $v_\M^2/2$ &-&$4.9\times10^{-9} $&$4.9\times10^{-9} $&  $1.2\times10^{-19} $\\
\hline
 $w_\mathrm{0M}$&Sun &$9.9\times10^{-9}$&$9.9\times10^{-9}$&$1.6\times10^{-19} $\\
		&Mercury &$ 1.7\times10^{-15}$&$ 1.7\times10^{-15}$ &$-6.9\times10^{-21}$\\ 
		&Venus &$ 2.9\times10^{-14}$ &$ 2.9\times10^{-14}$ &$3.1\times10^{-25}$\\
		&Earth&$1.2\times10^{-11} $&$1.2\times10^{-11} $&$-1.5\times10^{-21}$\\
		&Mars System& $2.4\times10^{-15} $& $2.4\times10^{-15} $&$-3.4\times10^{-23}$\\ 
		&Jupiter System& $ 1.8\times10^{-12} $& $ 1.8\times10^{-12} $ &$1.1\times10^{-20}$\\ 
		&Saturn System& $ 3.0\times10^{-13} $ & $ 3.0\times10^{-13} $&$2.8\times10^{-21}$\\ 
		&Uranus System  &$ 2.2\times10^{-14}$&$ 2.2\times10^{-14}$&$-2.9\times10^{-20}$\\
		&Neptune System & $ 1.7\times10^{-14}$ & $ 1.7\times10^{-14}$&$3.2\times10^{-21}$\\ 
		&Pluto System  &$1.8\times10^{-18} $&$1.8\times10^{-18} $&$2.8\times10^{-21}$\\ 
		&343 main belt asteroids&$4.6\times10^{-18}$&$4.7\times10^{-18}$&$-5.8\times10^{-20}$\\
		&30 \kbo s and a ring&0&$1.8\times10^{-17}$&$-1.8\times10^{-17}$\\
\hline
$w_{l\mathrm{M}}$	&$J_2$ of the Sun  &$2.2\times10^{-20}$& $2.3\times10^{-20}$&$-8.9\times10^{-22}$\\
		& $J_2$ of the Earth& $1.3\times10^{-18}$& $1.3\times10^{-18}$&$6.7\times10^{-26}$\\
\hline
post-Newtonian & all& $1.1\times10^{-16}$ & $1.1\times10^{-16}$	&$-8.4\times10^{-26}$\\
\hline
\end{tabular}
\end{table*}

We also estimate the contributions to $L_{C}^{\M}$ from different sources in \texttt{LTE440}, as shown in Table \ref{tab:LC_i}.
We find that the square of lunar velocity and the gravitational potential of the Sun at the Moon play most significant roles in $L_{C}^{\M}$, 
with the contribution reaching about $5\times 10^{-9}$ and $1\times10^{-8}$ respectively.
These contributions are at least two and three orders of magnitude bigger than those of Earth and Jupiter system,
which are the third and forth contributors in $L_{C}^{\M}$.
Despite remoteness, the objects and the ring in the Kuiper belt, which are new gravitational ingredients for DE440, contribute about $2\times 10^{-17}$ in $L_{C}^{\M}$, more than 3 times bigger than the overall of closer main belt asteroids, 
due to the much larger total mass of the Kuiper belt objects and the ring.
While we take the dynamical form factors $J_2$ of the Earth and the Sun into account for completeness,
the former climbs barely to about $1\times 10^{-18}$ and the latter is even smaller by two orders of magnitude.
Total contribution of all of the gravitational bodies at the post-Newtonian order $\mathcal{O}(c^{-4})$ is about $1\times 10^{-16}$, mainly coming from the Moon's velocity and the Sun's potential.

Using Fourier transform on \texttt{LTE440} after its linear trend removed, we find that there are 13 periodic terms with amplitudes larger than 1 $\muup$s.
The amplitudes $A_i$, periods $T_i$, and phases $\phi_i$ of these terms $A_i\sin(2\pi{T_i}^{-1}(t-\mathrm{J}2000.0)+\phi_i)$ are listed in Table \ref{tab:LTE440_Per}. 
The arguments of each term are derived from their periods, which indicate the contributing sources of each term. 
The first periodic variation with the largest amplitude of 1.6 ms and the period of about 365.26 day is caused by the orbital motion of the Earth-Moon barycenter around Sun. 
The second one with the amplitude of 126 $\muup$s and the period of 29.53 day comes from the orbital motion of the Moon around the Earth. 
The sources of the third and forth term are the motion of the Moon relative to the Jupiter and around Sun, respectively.
All the others variation with amplitudes below 10 $\muup$s are associated with the motion of the Moon, Venus, Earth, Jupiter, Saturn and Uranus.
We notice that the accuracy of these periodic terms is very limited by the Fourier transform and the figures in Table \ref{tab:LTE440_Per} should be taken as very preliminary results.
In our following works, we might use more sophisticated approaches to harmonically decompose \texttt{LTE440}, such as by the method of \citet{Harada2003AJ126.2557}.

\subsection{Estimated accuracy}

Since a \lte\ has a much more significant secular drift plus tiny variations,
we suppose that the error of the secular drift rate might dominate the accuracy of the \lte\ for the long time span.
According to Eq.~\eqref{eq:LTE}, this drift rate is inversely proportional to the distances between the Moon and other gravitational bodies. 
Taking the post-fit residuals of the ranging data adopted in DE440/441 \citep[see Section 5.1 in ][]{Park2021AJ161.105} as order-of-magnitude estimation of the distances error, 
we might preliminarily estimate that the error of the drift rate is about $7\times10^{-20}$, 
and we might find that the accuracy of \texttt{LTE440} is better than 0.15 ns within the year of 2050. We think that a more robust estimate of accuracy might have to wait for next version of JPL ephemeris.

\subsection{Comparison of \texttt{LTE430} and \texttt{LTE440}}
For better understandings of our products of \ltes, we make a comparison of \texttt{LTE430} and \texttt{LTE440}. 
We assume that differences between them might be mainly caused by the planetary and lunar data we used,
instead of our numerical integration scheme,
since this scheme has been tested to be accurate enough (see Sect. \ref{sec:alg} for details).
We find that, like \texttt{LTE440}, \texttt{LTE430} also has a more significant secular drift with indistinct periodic variations. 
For \texttt{LTE430}, its $L_{C}^{\M}$ indicating  secular drift rate between TCL and TCB is estimated as 
\begin{equation}
\label{eq:L_C^M_430}
 L_{C,430}^{\M}= 1.482\,536\,214\,9\times10^{-8}\pm1.0\times10^{-17},
\end{equation}
and it also has 13 periodic terms with amplitudes larger than 1 $\muup$s listed in Table \ref{tab:LTE440_Per}.

\begin{figure}[t!]
  \centering
  \includegraphics[width=\hsize]{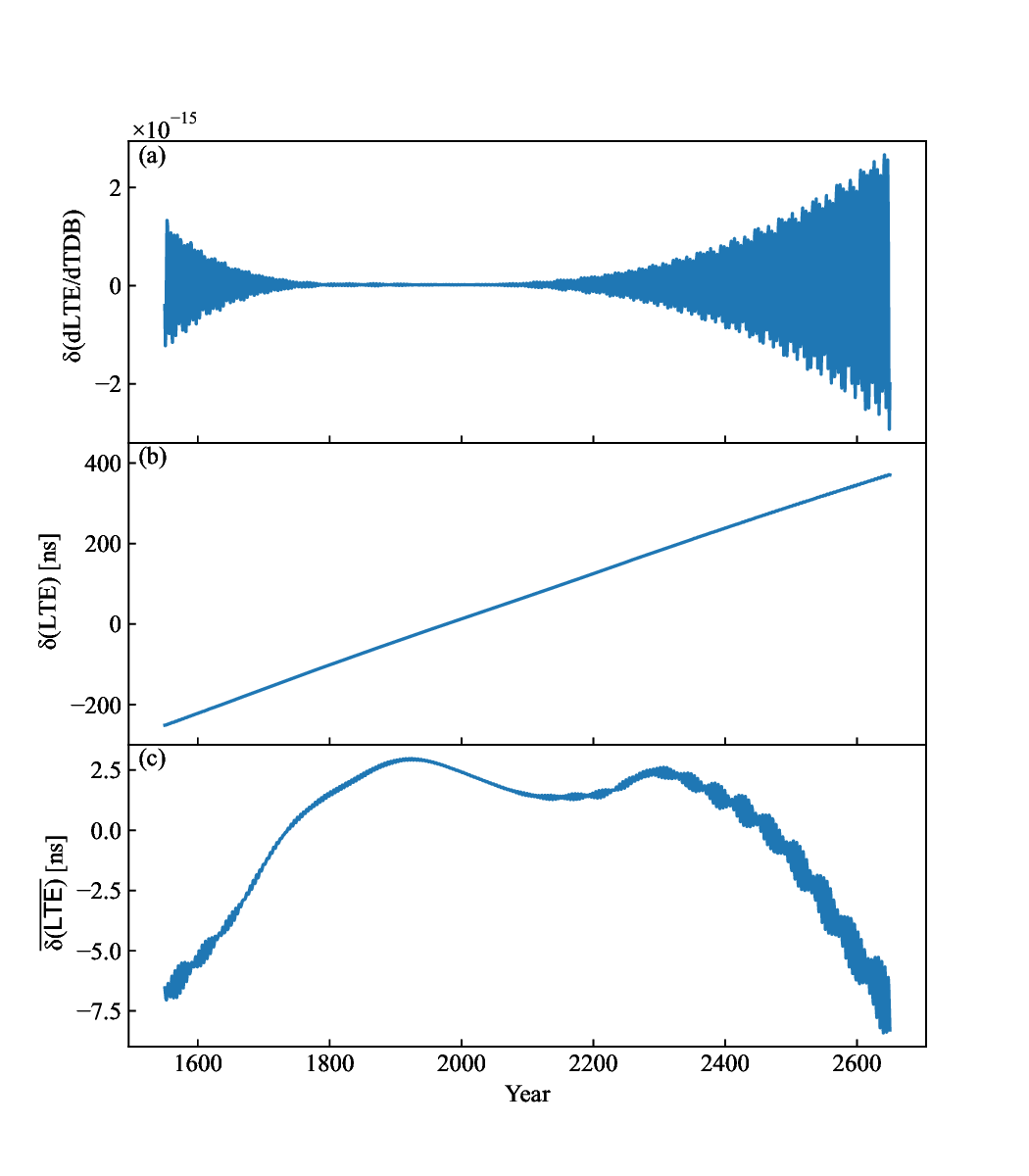}
   \caption{Comparison of LTE$_{430}$ and LTE$_{440}$. (a) Difference between their derivatives: $\delta(\mathrm{dLTE}/\mathrm{dTDB})=\mathrm{dLTE_{430}}/\mathrm{dTDB}-\mathrm{dLTE_{440}}/\mathrm{dTDB}$. (b) Difference between LTE$_{430}$ and LTE$_{440}$: $\delta\mathrm{LTE}=\mathrm{LTE_{430}}-\mathrm{LTE_{440}}$. (c) Detrended difference between LTE$_{430}$ and LTE$_{440}$: $ \overline{\delta(\mathrm{LTE})}=\delta(\mathrm{LTE})-\mathrm{linear\ drift}$.}
  \label{fig:LTE430mLTE440_comparison}
\end{figure}

The difference between the integrands of LTE$_{430}$ and LTE$_{440}$ is at the level of $2\times10^{-17}$ around the year 2000 and increases to about $2\times 10^{-15}$ over next 600 years, as shown in Fig.~\ref{fig:LTE430mLTE440_comparison}(a).
After integration, the resulting LTE$_{430}$ deviates from LTE$_{440}$ up to 13 ns around the year 2000 and their difference climbs up to about 350 ns around the year 2600, see Fig.~\ref{fig:LTE430mLTE440_comparison}(b),
because these two \ltes\ have slightly different $L_C^{\M}$ that (see Table~\ref{tab:L_C^M} for a summary)
\begin{equation}
\label{eq:L_C^M_430-L_C^M_440}
 L_{C,430}^{\M}-L_{C,440}^{\M}= -1.8\times10^{-17}.
\end{equation}
To pin down the cause of such a tiny but nonzero difference, 
we decompose $L_C^{\M}$ of \texttt{LTE430} and \texttt{LTE440} into its each contributing sources (see Table~\ref{tab:LC_i} for details).
We think that this difference is a natural consequence of 30 Kuiper belt objects and a ring in \texttt{LTE440}, which are absent in \texttt{LTE430},
due to the fact that their contribution matches the figure very well.
In fact, these remote \kbo s have almost 4 times more overall impact on $L_C^{\M}$ than 343 closer main belt asteroids.
For the periodic variations in \texttt{LTE430} and \texttt{LTE440},
the differences between their amplitudes with identical periods are no more than 12 ps, 
while the differences between their phases are less than $10^{-5}$ rad  (see Table \ref{tab:LTE440_Per}).
After the removal of the linear trend in Fig.~\ref{fig:LTE430mLTE440_comparison}(b), we reveal that the detrended difference between LTE$_{430}$ and LTE$_{440}$ ranging from -7.5 ns to 2.5 ns over the year span from 1550 to 2650 and it seems to have a bimodal feature and no distinct long periodic variations, as Fig.~\ref{fig:LTE430mLTE440_comparison}(c) demonstrated.

\begin{figure}[t!]
  \centering
\includegraphics[width=\hsize]{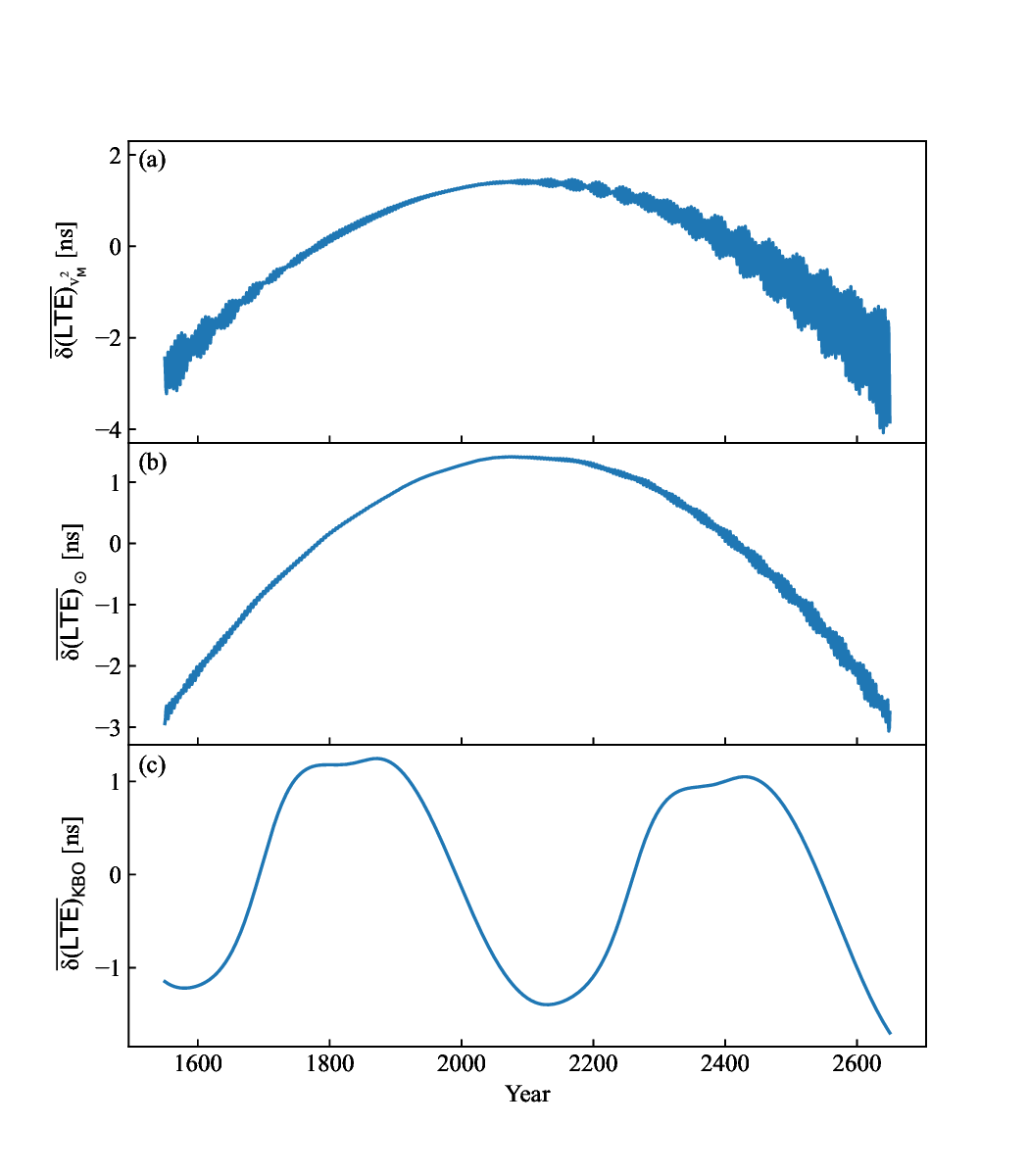}
   \caption{Detrended differences between mainly contributing sources of LTE$_{430}$ and LTE$_{440}$: (a) the square of Moon's velocity $\overline{\delta(\mathrm{LTE})}_{v^2_\M}$,  (b) the gravitational potential of the Sun at the Moon $\overline{\delta(\mathrm{LTE})}_\odot$, and (c) the gravitational potentials of the \kbo s and ring at the Moon $\overline{ \delta(\mathrm{LTE})}_\mathrm{KBO}$.}
  \label{fig:detrended(LTE430mLTE440)_comparison}
\end{figure}

In order to decipher the possible causes of such a detrended difference, we calculate the detrended contributions of each sources in the difference between \texttt{LTE430} and \texttt{LTE440}.
We find that it might trace back to three main sources: the Moon's velocity, the Sun's gravitational potential at the Moon and the gravitational potentials of the \kbo s at the Moon, whose contributions to the detrended LTE$_{430}-$LTE$_{440}$ are shown in Fig.~\ref{fig:detrended(LTE430mLTE440)_comparison}(a), (b) and (c) and range from -4 ns to 1.5 ns, from -3 ns to 1 ns and from -1.5 ns to 1 ns, respectively.
The contribution of the \kbo s and ring in the \texttt{LTE440} shows an oscillation with a period about 600 years, which could explain the bimodal shape in Fig.~\ref{fig:LTE430mLTE440_comparison}(c).
After adding them up, we could reproduce the result of detrended LTE$_{430}-$LTE$_{440}$.

In the works of \citet{Fukushima1995A&A294.895} and \citet{Irwin1999A&A348.642},
it was found that the mass-dependent errors from the planetary and lunar ephemerides are among the main reasons to make different versions of Earth time ephemerides deviate from each other, such as TE200 and TE405.
However, since current planetary and lunar ephemerides have been considerably improved with much more accurate and precise masses for the gravitational bodies than those adopted in \citet{Fukushima1995A&A294.895} and \citet{Irwin1999A&A348.642},
we suppose that this situation may no longer be true for our \ltes.
We estimate the effects of the mass-dependent errors between \texttt{LTE430} and \texttt{LTE440} on $L_{C}^{\M}$ and the amplitudes of the periodic variation with the analytic method used in \citet{Fukushima1995A&A294.895} (see Table \ref{tab:Mass-dependent error}).
We find that the differences caused by these mass-dependent errors are much less than the deviations we revealed (see Fig.~\ref{fig:LTE430mLTE440_comparison}) and could not be the main causes of them. 

\begin{table*}
\centering
  \caption{Results of Fourier transformation analysis for the periodic terms in LTE$_{430}$ and LTE$_{440}$ in the form of $A_i\sin[2\pi{T_i}^{-1}(t-\mathrm{J}2000.0)+\phi_i]$ with the amplitude $A_i>1$ $\muup$s, period $T_i$, and phase $\phi_i$.}
\label{tab:LTE440_Per} 
\begin{tabular}{rrrrrrc}
\hline\hline
  \multirow{2}{*}{Index $i$} & \multirow{2}{*}{Period $T_i$ [day]}& \multicolumn{2}{c}{LTE$_{430}$} & \multicolumn{2}{c}{LTE$_{440}$}  &\multirow{2}{*}{Arguments\tablefootmark{a}}\\
  &  & Amplitude $A_i$ [$\mu$s] & Phase $\phi_i$ [rad] & Amplitude $A_i$ [$\mu$s] & Phase $\phi_i$ [rad]&  \\
\hline
  1  &   365.26590909  & 1651.36355204 & 3.10895171 & 1651.36355077 & 3.10895165&  E\\
  2  &    29.53053800  &  126.30811930 & 5.18472342 &  126.30813184 & 5.18472464&  D\\
  3  &   398.99950348  &   19.37467866 & 1.33855840 &   19.37467715 & 1.33855843&  E$-$J\\
  4  &   182.63295455  &   13.70088726 & 3.07602307 &   13.70088760 & 3.07602294&  2E\\
  5  &   411.67264344  &    7.47520620 & 3.32446376 &    7.47520418 & 3.32446352&  D$-$L \\
  6  &  4320.34946237  &    4.24396687 & 3.43186307 &    4.24397312 & 3.43186281&  J \\
  7  &   377.97977422  &    3.76051419 & 0.92358653 &    3.76051430 & 0.92358639&  E$-$S\\
  8  &    14.25402654  &    2.93367408 & 1.09317021 &    2.93368121 & 1.09317212&  D$+$L\\
  9  &   369.63431463  &    2.67753135 & 1.51225313 &    2.67752983 & 1.51225314&  E$-$U\\
  10 &    32.12797857  &    2.36687830 & 5.21748673 &    2.36687890 & 5.21748801&  E$-$D \\
  11 & 10859.25675676  &    1.85819562 & 2.56842880 &    1.85820098 & 2.56843762&  S\\
  12 &   584.00072674  &    1.09742581 & 4.67635088 &    1.09742615 & 4.67635157&  V$-$E\\
  13 &   292.00036337  &    1.08850649 & 2.99249018 &    1.08850698 & 2.99248981&  2V$-$2E\\
\hline
\end{tabular}
\tablefoot{\tablefoottext{a}{The argument of each term is a combination of the mean longitudes of the solar system planets and the Delaunay elements of the Moon. Here ``V, E, J, S, U'' represent the mean longitudes of Venus, Earth-Moon barycenter, Jupiter, Saturn and Uranus respectively, ``D'' is the mean longitudes difference between the Sun and the Moon, and ``L'' is Moon's mean anomaly.}}
\end{table*}

\begin{table*}
	\centering
	\caption{Differences of $L_{C}^{\M}$ and amplitudes of the periodic variations caused by the mass-dependent errors between LTE$_{430}$ and LTE$_{440}$.}
	\begin{tabular}{llrlr}
		\hline\hline
		Contributing body $i$&$L_{C,i}^{\M}$&$\Delta M_i/M_i\times L_{C,i}^{\M}$&$A_i$[s]&$|\Delta M_i/M_i|\times A_i$[s]\\
		\hline
		Sun &$ 1.5\times10^{-8}$& $ 7.4\times10^{-20}$&$ 1.6\times10^{-3}$&$ 7.9\times10^{-15}$\\
		Mercury &$ 1.7\times10^{-15}$&$ -6.8\times10^{-21}$&$ 1.9\times10^{-9}$& $ 7.6\times10^{-15}$ \\ 
		Venus  & $ 2.9\times10^{-14}$&$ 4.1\times10^{-30}$&$ 2.0\times10^{-7}$& $ 2.8\times10^{-23}$\\
		Earth&$ 1.2\times10^{-11}$&$ 1.5\times10^{-21}$ &$ 2.9\times10^{-7}$&$ 3.7\times10^{-17}$\\
		Mars System& $ 2.4\times10^{-15}$ &$ -3.3\times10^{-23}$&$ 5.0\times10^{-8}$&$ 7.0\times10^{-16}$ \\ 
		Jupiter System&$ 1.8\times10^{-12}$ &$ 9.9\times10^{-21}$ &$ 7.5\times10^{-6}$& $ 4.1\times10^{-14}$ \\ 
		Saturn System & $ 3.0\times10^{-13}$ &$ 2.8\times10^{-21}$ &$ 2.5\times10^{-6}$&$ 2.4\times10^{-14}$  \\ 
		Uranus System  & $ 2.2\times10^{-14}$&$ -3.0\times10^{-20}$ &$ 4.5\times10^{-7}$&$ 6.0\times10^{-13}$\\
		Neptune System  & $ 1.7\times10^{-14}$&$ -9.3\times10^{-28}$ &$ 1.2\times10^{-7}$&$ 6.6\times10^{-21}$\\ 
		Pluto System  & $ 1.8\times10^{-18}$& $ 2.8\times10^{-21}$ &$ 5.8\times10^{-10}$&$ 8.9\times10^{-13}$\\ 
		\hline
		Total&$ 1.5\times10^{-8}$& $ 5.4\times10^{-20}$& --&-- \\
		\hline
	\end{tabular}
	\tablefoot{$L_{C,i}^{\M}$ is derived from Table \ref{tab:LC_i} and $A_i$ is estimated from the periodic variation of each contributing body in LTE$_{440}$. $L_{C,i}^{\M}$ and $A_i$ of the Sun are assumed to be the sum of the contribution by the square of lunar velocity and the gravitational potential of the Sun at the Moon. $\Delta M_i/M_i$ means the relative mass difference of body $i$ between DE430 and DE440. }
	\label{tab:Mass-dependent error}
\end{table*}

Therefore, we think that the existence of the \kbo s and ring in DE440 might cause the long-term deviation between \texttt{LTE430} and \texttt{LTE440},
while the improvement of Moon's position and velocity as well as the \kbo s in DE440 could explain the remainder of detrended LTE$_{430}-$LTE$_{440}$.
Except these factors, the rest parts of \texttt{LTE430} and \texttt{LTE440} have very good consistency overall.


\subsection{\texorpdfstring{Correction to $L_{C}^{\M}$}{Correction to LCM}}


In the works of \citet{Irwin1999A&A348.642} on the numerical Earth \tes,
these authors did not recommend to determine the secular drift rate of TCG with respect to TCB by numerical approaches alone,
because this procedure strongly depends on the epoch range chosen for the integration.
Instead, they thought a much better approach to determine such a rate was to use an analytic series approximation for the Earth \te\ and then find the coefficient of the linear term as the drift rate.
This series approach takes the advantage of the fact that an analytic series usually has a much longer time interval than the numerical ephemerides and one might recognize the periodic and secular terms without any ambiguity. 
Nevertheless, considering the less accuracy of the series than the one of the numerical ephemeris, 
these authors proposed a hybrid approach that applies a linear least-squares fit to the residuals between a series with the linear drift removed and a numerical time ephemeris with a preliminary value of the drift rate removed as well,
which can give a correction to the rate obtained by the numerical time ephemeris.

We think that this statement is also true for the determination of  $L_{C}^{\M}$ in the \lte.
However, we are facing a frustrating fact that 
any analytic series for the \lte\ as sophisticated as those of Earth \tes\  \citep{Hirayama1987TSCM.75,Fairhead1990A&A229.240} is currently unavailable,
and we expect that it might take a quite long time to achieve such an ambitious goal for making one.
Therefore, in order to bypass this situation and find a timely way,
we propose a mixed numerical approach to calibrate $L_{C}^{\M}$ of a short \lte\ by a much longer \lte.
Since \texttt{LTE441} has a much longer time span than \texttt{LTE440},
it is used to do the calibration of $L_{C,440}^{\M}$.
We apply a linear least-squares fit of the residuals between LTE$_{441}$ and LTE$_{440}$, both of which are linearly detrended.
This fit gives a correction to the numerical value to $L_{C,440}^{\M}$
that is independent of the numerical value to $L_{C,441}^{\M}$.
Therefore, we find the calibrated $L_{C,440}^{\M}$ as
\begin{equation}
\label{eq:L_C^M_440|441}
 L_{C,440\,|\,441}^{\M}= 1.482\,536\,221\,7\times10^{-8},
\end{equation}
which is tinily bigger than $L_{C,440}^{\M}$.
It can slow down the drift between TCL and TCB 
by removing the long-term periodic contribution from the secular drift rate, see Eq.~\eqref{eq:<dTCLdTCB>}.

\section{Conclusions}
\label{sec:con}

We present a numerical \lte\ package \texttt{LTE440}, 
which contains the data files of the evaluated relativistic time-dilation integral
and software functions to calculate transformations among TCL, TCB and TDB.
We compute the integral by using 10th-order Romberg scheme with quantities from JPL ephemeris DE440 and produce the data files by 13th-degree Chebyshev polynomials for interpolation.
The accuracy of \texttt{LTE440} is estimated to be better than 0.15 ns in the lunar time ephemeris and $7 \times 10^{-20}$ in its secular drift rate within the year of 2050.
We believe that these levels of accuracy might satisfy most of current needs.

We also make two auxiliary lunar time ephemeris packages \texttt{LTE430} and \texttt{LTE441} based on DE430 and DE441 for performance assessment. 
After a detailed comparison of \texttt{LTE430} and \texttt{LTE440},
we find that the most parts of them have very good consistency overall,
while the existence of the Kuiper belt objects and the improvement of Moon's motion in DE440 might cause differences between them.
In order to separate secular drift from periodic variations with a better unambiguity,
we calibrate the long-term change rate of TCL and TCB in \texttt{LTE440} with the help of \texttt{LTE441}. 

We believe that an external comparison of \texttt{LTE440} with other \ltes\ based on INPOP21a and EMP2021 would be helpful to know their performances and characteristics and provide more insights into them,
and we will leave it as our next moves.

\begin{acknowledgements}
      This work is funded by
the Strategic Priority Research Program on Space Science of the Chinese Academy of Sciences (XDA300103000, XDA30040000, XDA30030000 and XDA0350300) and
 the National Natural Science Foundation of China (Grants No. 62394350, No. 62394351 and No. 12273116).
\end{acknowledgements}

%

\bibliographystyle{aa}
\bibliography{refs}

\begin{appendix}




\begin{table*}
  \section{Key information of DE430, DE440 and DE441}
   \caption{Comparison of DE430, DE440 and DE441\newline}
   \label{tab:dynmod}
   \centering
  \begin{tabular}{p{0.19\linewidth}p{0.24\linewidth}p{0.24\linewidth}p{0.23\linewidth}}
    \hline\hline 
     & DE430 & DE440 & DE441 \\
    \hline 
    Time Spans
    & 1550 to 2650
    & 1550 to 2650 
    & $-$13 200 to $+$17 191\\
    \hline 
    Gravitational Bodies
    & Sun, 8 planets, Moon\par Pluto \par 343 main belt asteroids
    & Sun, 8 planets, Moon\par Pluto \par 343 main belt asteroids\par 30 \kbo s and a ring
    & same as DE440\\
    \hline 
    Point-mass Interaction
    &all gravitational bodies 
    & all gravitational bodies 
    & same as DE440\\
    \hline
    Nonspherical Interaction
    & up to the second-degree zonal harmonic of the Sun\par up to the fourth-degree zonal harmonic of the Earth\par up to the sixth-degree zonal and tesseral harmonic of the Moon 
    &up to the second-degree zonal harmonic of the Sun\par up to the fifth-degree zonal harmonic of the Earth\par up to the sixth-degree zonal and tesseral harmonic of the Moon
    &same as DE440\\ 
    \hline
    Other Effects
    &effects on the Moon's motion by Earth tides
    &effects on the Moon's motion by Earth tides\par Lense-Thirring effect from the Sun's angular momentum\par solar radiation pressure force on the Earth and Moon
    &same as DE440, but without damping between the lunar liquid core and solid mantle\\
    \hline 
  \end{tabular}
\end{table*}

\end{appendix}
\end{document}